# PRACTICAL DATA CORRELATION OF FLASHPOINTS OF BINARY MIXTURES BY A RECIPROCAL FUNCTION: the concept and numerical examples

By

*Mariana HRISTOVA[1], Dimitar DAMGALIEV[2] , Jordan HRISTOV[3*]*

1-Department of General Chemical Technology, 2-Department of Automation, 3-Department of Chemical Engineering, , University of Chemical Technology and Metallurgy, Sofia, Bulgaria.

E-mail: jordan.hristov@mail.bg

*Simple data correlation of flashpoint data of binary mixture has been developed on a basic of rational reciprocal function. The new approximation requires has only two coefficients and needs the flashpoint temperature of the pure flammable component to be known. The approximation has been tested by literature data concerning aqueous-alcohol solution and compared to calculations performed by several thermodynamic models predicting flashpoint temperatures. The suggested approximation provides accuracy comparable and to some extent better than that of the thermodynamic methods.*

## Introduction

The flash points (FP) of flammable or combustible liquids are data required to establish the fire and explosion hazards and classifications of materials according to the classes defined in each particular regulation [1, 2]. This, therefore, requires knowledge related to both the principles of combustion and the fluid phase equilibria. The flash point is defined as the “lowest temperature (corrected to $101.3\,kPa$ ) at which the vapors of a specimen ignite , under specified conditions of a test” [1], by application of an external ignition source, and therefore the lower explosion limit exceeds the flash point [3]. The flash points are almost constant characteristics of materials tested but the published values vary because they strongly depend on the design of the testing device. Usually, the closed-cup method [4] is used because the results tend to be on the safe side, while the open-cup measurements [4] are not reliable, to some extent, due to a systematic errors caused by volatile compounds escape from the measuring equipment. A particular flash point can therefore only be defined in terms of a particular standardized test method.

The existing estimation methods, especially for closed cup flash points [5-9] are based on iterative calculations and use combinations of a) Dalton’s and , Raoult’s laws for ideal solutions [10]; b) corrected Raoult’s law for non-ideal solutions [10], the Antoine equation [11], and the Le Chatelier’s rule [12]. Many FP data published in the literature are based on experiments, thus incorporating either experimental or systematic errors. A statistical analysis; therefore, providing handy relationships is highly required.

The present communication addresses handy flash point approximation of binary aqueous mixtures by a reasonable relationship allowing predicting easily $T_{FP}$ when only the concentration of the flammable component ($x_2$) is known. We address water-alcohol mixtures (data taken from [13] as good examples with non-ideal behaviours, allowing to calculate $T_{FP}$ by different thermodynamic models and iterative calculations (by Matlab), as parallel prediction procedures [14].

## Approximations developed

Commonly, a 3rd order polynomial correlations for FP is used [17] to fit particular sets of experimental data, i.e.

$$FP = b_0 + b_1 x + b_2 x^2 + b_3 x^3 \qquad (1)$$

where $x = x_2$, the molar fraction of the flammable component (FC) of the water ($x_1$)-FC ($x_2$) mixture .

This type of relationships has 4 coefficients inherently affected by the uncertainty in the experiments and the regression procedure. The present works conceive an a approximating relationship from the family of the rational functions, the so-called reciprocal function, requiring an initial normalization of the experimental FP data by $T_{FP}$ of the pure flammable component is conceived, namely

$$y = \frac{1}{a+bx}, \quad y = T_{FP} / T_{FP(x=1.0)} \qquad (2)$$

In (2), we have $a+b \approx 1$ because at $x \to 1$, we have $y \to 1$.

## Numerical Experiments

Data correlations by equations developed (see Table 1) for some water-based binary mixtures on the basis of (1) and (2) were performed together with iterative calculations [16] based on thermodynamic models. The outcomes concerning some sample mixtures [14] are summarized in Tables 2-4. The data summarized indicate almost equal level off approximation (based on the absolute point-wise errors) of both the empirical approximations and the prediction of the thermodynamic models. The conceived reciprocal function fits the experimental data better than the 3rd polynomial expressions. Errors of comparable to those provided by the polynomial relationships were observed with water-iso-propanol mixtures (Table 5) only. Some special features and advantages of the suggested approximation functions are commented next.

## Comments

The equations developed have only goal: to fit the experimental data with a minimum error of approximations. In general, both type of equations used lead to almost equal errors of approximations within the range $0.1 \le x \le 1.0$ . The reciprocal approximation is more practical because only two coefficients $a$ and $b$, as well as , the $T_{FP}$ of the pure flammable component, have to be known. The polynomial approximation needs 4 coefficients but $T_{FP}$ is not needed for the calculations. However, the $T_{FP(x=1)}$ is a useful initial datum allowing to normalize the experimental data as $y = T_{FP} / T_{FP(x=1.0)}$ and control the adequacy of approximation taking into account that at $x \to 1$, we have $y \to 1$ and $a+b \approx 1$.

Table 1. Flash-point approximations developed

| **Binary mixture** | **Polynomial approximation** | **Reciprocal approximation** |
|---|---|---|
| Water-Methanol (x) | $FP = 56.7 - 148x + 187x^2 - 86.1x^3$<br>$R^2 = 0.995$; SSE=5.52 | $T_{FP}/T_{FP(x=1.0)} = (0.148 + 0.765x)^{-1}$<br>$R^2 = 0.995$; $\chi^2 = 0.00576$ |
| Water-Ethanol (x) | $FP = 40.2 - 86.1x + 122x^2 - 63x^3$<br>$R^2 = 0.983$; SSE=5.54 | $T_{FP}/T_{FP(x=1.0)} = (0.345 + 0.608x)^{-1}$<br>$R^2 = 0.964$; $\chi^2 = 0.00856$ |
| Water-Propanol (x) | $FP = 31.5 - 5.29 - 8.7x^2 + 5.36x^3$<br>$R^2 = 0.984$; SSE=1.12 | $T_{FP}/T_{FP(x=1.0)} = (0.705 + 0.290x)^{-1}$<br>$R^2 = 0.976$; $\chi^2 = 0.0004$ |
| Water-iso-Propanol (x) | $FP = 26.2 - 34.0x + 48.2x^2 - 27.6x^3$<br>$R^2 = 0.994$; SSE=0.53 | $T_{FP}/T_{FP(x=1.0)} = (0.526 + 0.413x)^{-1}$<br>$R^2 = 0.967$; $\chi^2 = 0.00208$ |

Table2 Mixture flash points of water (1)-methanol (2)

| Exp. Data | | Thermodynamic-based model predictions | | | | | | | | Approximations (present work) | | | |
|---|---|---|---|---|---|---|---|---|---|---|---|---|---|
| $x_2$ | $T_{FP}$ $^oC$ (exp) | $T_{FP}$ $^oC$ (M) | $\Delta_{ME}$ $^oC$ | $T_{FP}$ $^oC$ (vL) | $\Delta_{vLE}$ $^oC$ | $T_{FP}$ $^oC$ (W) | $\Delta_{WE}$ $^oC$ | $T_{FP}$ $^oC$ (Ideal) | $\Delta_{IE}$ $^oC$ | $T_{FP}$ $^oC$ appr. (P) | $\Delta_{poly}$ $^oC$ appr. (P) | $T_{FP}$ $^oC$ appr. (R) | $\Delta_{power}$ $^oC$ appr. (R) |
| 1.0 | 10.0 | | | | | | | | | 9.2 | -0.8 | 10.93 | 0.93 |
| 0.9 | 10.6 | 11.7 | 1.1 | 11.7 | 1.1 | 11.7 | 1.1 | 11.8 | 1.2 | 11.9 | 1.3 | 11.93 | -1.33 |
| 0.8 | 13.7 | 13.6 | -0.1 | 13.6 | -0.1 | 13.4 | -0.3 | 13.9 | 0.2 | 13.7 | 0.0 | 13.13 | -0.56 |
| 0.7 | 15.6 | 15.5 | -0.1 | 15.5 | -0.1 | 15.3 | -0.3 | 16.2 | 0.6 | 15.1 | -0.5 | 14.60 | -0.99 |
| 0.6 | 16.3 | 17.6 | 1.3 | 17.6 | 1.3 | 17.2 | 0.9 | 19.0 | 2.7 | 16.5 | 0.2 | 16.44 | 0.14 |
| 0.5 | 19.2 | 19.9 | 0.7 | 19.9 | 0.7 | 19.3 | 0.1 | 22.4 | 3.2 | 18.6 | -0.6 | 18.81 | -0.38 |
| 0.4 | 22.3 | 22.6 | 0.3 | 22.6 | 0.3 | 21.8 | -0.5 | 26.7 | 4.4 | 21.9 | -0.4 | 21.98 | -0.31 |
| 0.3 | 26.7 | 26.1 | -0.6 | 26.3 | -0.4 | 25.1 | -1.6 | 32.4 | 5.7 | 26.8 | 0.1 | 26.43 | -0.26 |
| 0.2 | 32.6 | 31.6 | -1.0 | 31.8 | -1.2 | 30.2 | -2.4 | 40.9 | 8.3 | 33.9 | 1.3 | 33.14 | 0.54 |
| 0.1 | 44.5 | 42.8 | -1.7 | 42.9 | -1.6 | 40.8 | -3.7 | 56.7 | 12.2 | 43.7 | -0.8 | 44.42 | -0.07 |
| Exp: from Liaw [14]; M- Margules [15] ;vL - van Laar [15]; I- ideal (Raoult's law) [11,15]; P- polynomial; R – reciprocal; $\Delta = (T_{predicted} - T_{experimental})$ | | | | | | | | | | | | | |

Table 3 Mixture flash points of water(1)-ethanol (2)

| Exp. Data | | Thermodynamic-based model predictions | | | | | | | | Approximations (present work) | | | |
|---|---|---|---|---|---|---|---|---|---|---|---|---|---|
| $x_2$ | $T_{FP}$ $^oC$ (exp) | $T_{FP}$ $^oC$ (M) | $\Delta_{ME}$ $^oC$ | $T_{FP}$ $^oC$ (vL) | $\Delta_{vLE}$ $^oC$ | $T_{FP}$ $^oC$ (W) | $\Delta_{WE}$ $^oC$ | $T_{FP}$ $^oC$ Ideal | $\Delta_{IE}$ $^oC$ | $T_{FP}$ $^oC$ appr. (P) | $\Delta_{poly}$ $^oC$ appr. (P) | $T_{FP}$ $^oC$ appr. (R) | $\Delta_{power}$ $^oC$ appr. (R) |
| 1.0 | 13.0 | | | | | | | | | 12.5 | -0.5 | 13.61 | 0.61 |
| 0.9 | 14.6 | 14.8 | 0.2 | 14.8 | 0.2 | 14.5 | -0.1 | 14.7 | 0.1 | 15.1 | 0.5 | 14.54 | -0.05 |
| 0.8 | 16.3 | 16.7 | 0.4 | 16.5 | 0.2 | 15.8 | -0.5 | 16.6 | 0.3 | 16.8 | 0.5 | 15.60 | -0.69 |
| 0.7 | 17.5 | 18.6 | 1.1 | 18.3 | 0.8 | 17.1 | -0.4 | 18.7 | 1.2 | 17.8 | 0.3 | 16.83 | -0.66 |
| 0.6 | 19.5 | 20.4 | 0.9 | 20.0 | 0.5 | 18.3 | -1.2 | 21.3 | 1.8 | 18.7 | -0.8 | 18.27 | -1.22 |
| 0.5 | 20.4 | 21.9 | 1.5 | 21.7 | 1.3 | 19.4 | -1.0 | 24.4 | 4.0 | 19.7 | -0.7 | 19.98 | -0.41 |
| 0.4 | 20.8 | 23.4 | 2.6 | 23.4 | 2.6 | 20.6 | -0.2 | 28.3 | 7.5 | 21.2 | -0.4 | 22.05 | 1.25 |
| 0.3 | 24.0 | 25.0 | 1.0 | 25.3 | 1.3 | 22.0 | -2.0 | 33.5 | 9.5 | 23.6 | -0.4 | 24.59 | 0.59 |
| 0.2 | 25.8 | 27.6 | 1.8 | 28.1 | 2.3 | 24.1 | -1.7 | 41.2 | 15.4 | 27.3 | 1.5 | 27.79 | 1.99 |
| 0.1 | 33.6 | 34.5 | 0.9 | 34.6 | 1.0 | 29.1 | -4.5 | 55.5 | 21.9 | 32.8 | -0.8 | 31.95 | -1.64 |

Table 4 Mixture flash point of water(1)-propanol (2)

| Exp. Data | | Thermodynamic-based model predictions | | | | | | | | Approximations (present work) | | | |
|---|---|---|---|---|---|---|---|---|---|---|---|---|---|
| $x_2$ | $T_{FP}$ $^oC$ (exp) | $T_{FP}$ $^oC$ (M) | $\Delta_{ME}$ $^oC$ | $T_{FP}$ $^oC$ (vL) | $\Delta_{vLE}$ $^oC$ | $T_{FP}$ $^oC$ (W) | $\Delta_{WE}$ $^oC$ | $T_{FP}$ $^oC$ (Ideal) | $\Delta_{IE}$ $^oC$ | $T_{FP}$ $^oC$ appr. (P) | $\Delta_{poly}$ $^oC$ appr. (P) | $T_{FP}$ $^oC$ appr. (R) | $\Delta_{power}$ $^oC$ appr. (R) |
| 1.0 | 23.0 | | | | | | | | | 22.7 | -0.3 | 23.08 | 0.08 |
| 0.9 | 23.4 | 24.7 | 1.3 | 24.5 | 1.1 | 24.3 | 0.9 | 24.6 | 1.2 | 23.6 | 0.2 | 23.77 | 0.37 |
| 0.8 | 23.9 | 26.6 | 2.7 | 26 | 2.1 | 25.3 | 1.4 | 26.3 | 2.6 | 24.5 | 0.6 | 24.51 | 0.61 |
| 0.7 | 26 | 28.5 | 2.5 | 27,4 | 1.4 | 26.1 | 0.1 | 28.4 | 2.4 | 25.4 | -0.6 | 25.29 | -0.70 |
| 0.6 | 26.3 | 30 | 3.7 | 28.8 | 2.5 | 26.7 | 0.4 | 30.8 | 4.5 | 26.4 | 0.1 | 26.13 | -0.16 |
| 0.5 | 27.2 | 30.8 | 3.6 | 30 | 2.8 | 27.3 | 0.1 | 33.7 | 6.5 | 27.3 | 0.1 | 27.02 | -0.17 |
| 0.4 | 28.1 | 30.9 | 2.8 | 30.8 | 2.7 | 27.8 | -0.3 | 37.4 | 9.3 | 28.2 | 0.1 | 27.97 | -0.12 |
| 0.3 | 29.6 | 30 | 0.4 | 31.1 | 1.5 | 28.3 | -1.3 | 42.3 | 12.7 | 29.1 | -0.5 | 29.00 | -0.59 |
| 0.2 | 29.7 | 28.8 | -0.9 | 30.8 | 1.1 | 28.8 | -0.9 | 49.6 | 19.9 | 30.0 | 0.3 | 30.10 | 0.40 |
| 0.1 | 31 | 29.6 | -1.4 | 31 | 0 | 30 | -1 | 63.2 | 32.2 | 31.0 | 0.0 | 31.29 | 0.29 |

Table 5 Mixture flash point of water(1) - iso-propanol (2)

| Exp. Data | | Thermodynamic-based model predictions | | | | | | | | Approximations (present work) | | | |
|---|---|---|---|---|---|---|---|---|---|---|---|---|---|
| $x_2$ | $T_{FP}$ $^oC$ (exp) | $T_{FP}$ $^oC$ (M) | $\Delta_{ME}$ $^oC$ | $T_{FP}$ $^oC$ (vL) | $\Delta_{vLE}$ $^oC$ | $T_{FP}$ $^oC$ (W) | $\Delta_{WE}$ $^oC$ | $T_{FP}$ $^oC$ (Ideal) | $\Delta_{IE}$ $^oC$ | $T_{FP}$ $^oC$ appr. (P) | $\Delta_{poly}$ $^oC$ appr. (P) | $T_{FP}$ $^oC$ appr. (R) | $\Delta_{power}$ $^oC$ appr. (R) |
| 1.0 | 13.0 | | | | | | | | | 12.8 | -0.2 | 14.89 | 1.89 |
| 0.9 | 14.3 | 14.6 | 0.3 | 14.5 | 0.2 | 14.4 | 0.1 | 14.6 | 0.3 | 14.5 | 0.2 | 15.5 | 1.28 |
| 0.8 | 15.6 | 16.3 | 0.7 | 16 | 0.4 | 15.5 | -0.1 | 16.4 | 0.8 | 15.7 | 0.1 | 16.33 | 0.73 |
| 0.7 | 16.5 | 17.9 | 1.4 | 17.4 | 0.9 | 16.5 | 0 | 18.4 | 1.9 | 16.5 | 0.0 | 17.16 | 0.66 |
| 0.6 | 17.3 | 19.3 | 2 | 18.8 | 1.5 | 17.4 | 0.1 | 20.9 | 3.6 | 17.2 | -0.1 | 18.08 | 0.78 |
| 0.5 | 18 | 20.1 | 2.1 | 20 | 2 | 18.2 | 0.2 | 23.8 | 5.8 | 17.8 | -0.2 | 19.10 | 1.10 |
| 0.4 | 18.8 | 20.5 | 1.7 | 21 | 1.2 | 18.9 | 0.1 | 27.5 | 8.7 | 18.5 | -0.3 | 20.24 | 1.44 |
| 0.3 | 19.3 | 20.5 | 1.2 | 21.8 | 2.5 | 19.6 | 0.3 | 32.4 | 13.1 | 19.6 | 0.3 | 21.53 | 2.23 |
| 0.2 | 20.7 | 20.4 | -0.3 | 22.6 | 1.9 | 20.5 | -0.2 | 39.7 | 19 | 21.1 | 0.4 | 22.99 | 2.29 |
| 0.1 | 23.5 | 23 | -0.5 | 25.2 | 1.7 | 22.4 | -1.1 | 53.1 | 29.6 | 23.2 | -0.3 | 24.67 | 1.17 |